\documentclass[a4paper]{svmult}
\usepackage{times}
\usepackage[dvips]{graphicx,epsfig}
\usepackage{amsmath,amsfonts,amssymb}
\usepackage{cite,url}

\begin{document}

\title*{Exploring Effective Three-body Forces}
\author{\underline{{Alexander Volya}}}
\titlerunning{Exploring Effective Three-body Forces}
\authorrunning{{A.~Volya}}
\toctitle{Exploring Effective Three-body Forces}
\tocauthor{A.~Volya}
\institute{Department of Physics, Florida State University, Tallahassee, FL
32306-4350, USA}
%\date{\today}
\maketitle
\begin{abstract}
Topics related to the construction, phenomenological determination, and
effects of the effective three-body forces within the  
traditional nuclear shell model approach are discussed.  The
manifestations of the three-body forces in realistic nuclei in the
$0f_{7/2}$ and $1s0d$ 
shell model valence spaces are explored. 
\end{abstract}
\section{Introduction}

In this work we investigate the role that three-body forces
play within the nuclear shell model (SM) approach. Establishment of the 
effective interaction parameters, study of hierarchy in strength from
single-particle (s.p.) to two-body, three-body, and beyond,
manifestations in energy spectra and transitions rates, comparison
with different traditional SM calculations, and overall assessment for
the need of beyond-two-body SM are the topics for this discussion. 
Previous works in this direction have
shown an improved description of nuclear spectra \cite{Eisenstein:1973,
Poves:1981, Hees:1989, Hees:1990,Volya:2008, Volya:arxiv} and the significance of three-body
monopole renormalizations \cite{Zuker:2003}.

The effective interaction Hamiltonian of rank $k$ is a sum \begin{equation}
H_{k}=\sum_{n=1}^{k}H^{(n)},\,\,\,{\rm where}\,\, H^{(n)}=\sum_{\alpha\beta}\sum_{L}V_{L}^{(n)}(\alpha\beta)\sum_{M=-L}^{L}\, T_{LM}^{(n)^{\dagger}}(\alpha)\, T_{LM}^{(n)}(\beta),\label{eq:Hcn}\end{equation}
is the $n$-body rotationally invariant component of the interaction. The $n$-particle creation
operators $T_{LM}^{(n)^{\dagger}}(\alpha)$ with the total angular momentum
$L$ and magnetic projection $M$ are normalized $\langle0|T_{L'M'}^{(n)}(\alpha')\, T_{LM}^{(n)^{\dagger}}(\alpha)|0\rangle=\delta_{\alpha\alpha'}\delta_{LL'}\delta_{MM'}$
and expressed through the s.p. creation operators as $T_{LM}^{(n)^{\dagger}}(\alpha)=\sum_{12\dots n}C_{12\dots n}^{LM}(\alpha)\, a_{1}^{\dagger}a_{2}^{\dagger}\dots a_{n}^{\dagger},$
where index $1$ labels a s.p. state. The choice of coefficients $C_{12\dots n}^{LM}(\alpha)$
that defines a full set of orthogonal operators $T_{LM}^{(n)}(\alpha)$
is generally not unique. For numerical work it is most convenient
to use a full set of orthogonal eigenstates $|n;LM\alpha\rangle=T_{LM}^{(n)^{\dagger}}(\alpha)|0\rangle$
of some $n$-particle system \cite{Volya:2008}. 
In the $m$-scheme
SM we generate states only for a particular value of the
total magnetic projection $M$, the remaining states are obtained
by the raising and lowering angular momentum operators. It is possible \cite{Volya:2008}, to select a
single reference two-body Hamiltonian which then can 
be used to define all many-body operators $T_{LM}^{(n)^{\dagger}}(\alpha)$
for $n>2.$ 
The traditional
SM Hamiltonian is  $H_{2}=H^{(1)}+H^{(2)},$ where the two-body
operators $T_{LM}^{(2)^{\dagger}}(\alpha)$ are determined with the
help of the Clebsch-Gordan coefficients.

\section{Manifestation of three-body forces in f$_{7/2}$-shell nuclei}

As a first example we present here a study of a single-$j$ $0f_{7/2}$
shell, related discussion may be found in Ref.~\cite{Volya:arxiv}. We consider two types of systems $N=28$ isotones starting
from $^{48}$Ca with protons filling the $0f_{7/2}$ shell and the
Z=20, $^{40-48}$Ca isotopes with valence neutrons. The states in these
systems that are identified by experiments with the $f_{7/2}$ valence
space are listed in Tab. \ref{tab:Experimental-data}. The 
$f_{7/2}$ shell is unique because of symmetries associated with the quasispin and
particle-hole conjugation 
\cite{Talmi:1957,Ginocchio:1963,McCullen:1964,Eisenstein:1973,Volya:2002PRC}. 
These symmetries are violated if interaction is beyond the two-body.

\begin{table}
\begin{center}
\begin{tabular}{|c|c||c|c|c||c|c|c|}
\hline 
\multicolumn{2}{|c||}{} & \multicolumn{3}{c||}{$N=28$} & \multicolumn{3}{c|}{$Z=20$}\tabularnewline
\hline 
spin & $\nu$ & name & Binding & 3B$f_{7/2}$ & name & Binding & 3B$f_{7/2}$\tabularnewline
\hline
\hline 
0 & 0 & $^{48}$Ca & 0 & 0 & $^{40}$Ca & 0 & 0\tabularnewline
\hline
\hline 
7/2 & 1 & $^{49}$Sc & 9.626 & 9.753  & $^{41}$Ca & 8.360 & 8.4870 \tabularnewline
\hline
\hline 
0 & 0 & $^{50}$Ti & 21.787 & 21.713  & $^{42}$Ca & 19.843 & 19.837 \tabularnewline
\hline 
2 & 2 & 1.554 & 20.233 & 20.168  & 1.525 & 18.319 & 18.314 \tabularnewline
\hline 
4 & 2 & 2.675 & 19.112 & 19.158  & 2.752 & 17.091 & 17.172 \tabularnewline
\hline 
6 & 2 & 3.199 & 18.588 & 18.657  & 3.189 & 16.654 & 16.647 \tabularnewline
\hline
\hline 
7/2 & 1 & $^{51}$V & 29.851 & 29.954  & $^{43}$Ca & 27.776 & 27.908 \tabularnewline
\hline 
5/2 & 3 & 0.320 & 29.531 & 29.590  & .373 & 27.404 & 27.630\tabularnewline
\hline 
3/2 & 3 & 0.929 & 28.922 & 28.992  & .593 & 27.183 & 27.349\tabularnewline
\hline 
11/2 & 3 & 1.609 & 28.241 & 28.165  & 1.678 & 26.099 & 26.128 \tabularnewline
\hline 
9/2 & 3 & 1.813 & 28.037 & 28.034  & 2.094 & 25.682 & 25.747 \tabularnewline
\hline 
15/2 & 3 & 2.700 & 27.151 & 27.106  & 2.754 & 25.022 & 24.862 \tabularnewline
\hline
\hline 
0 & 0 & $^{52}$Cr & 40.355 & 40.292  & $^{44}$Ca & 38.908 & 38.736 \tabularnewline
\hline 
2 & $2^{*}$ & 1.434 & 38.921 & 38.813  & 1.157 & 37.751 & 37.509 \tabularnewline
\hline 
4 & $4^{*}$ & 2.370 & 37.986 & 38.002  & 2.283 & 36.625 & 36.570 \tabularnewline
\hline 
4 & $2^{*}$ & 2.768 & 37.587 & 37.643  & 3.044 & 35.864 & 36.009 \tabularnewline
\hline 
2 & $4^{*}$ & 2.965 & 37.390 & 37.183  & 2.657 & 36.252 & 35.741 \tabularnewline
\hline 
6 & 2 & 3.114 & 37.241 & 37.353  & 3.285 & 35.623 & 35.606 \tabularnewline
\hline 
5 & 4 & 3.616 & 36.739 & 36.789  & - & - & 35.180 \tabularnewline
\hline 
8 & 4 & 4.750 & 35.605 & 35.445  & (5.088) & (33.821) & 33.520 \tabularnewline
\hline
\hline 
7/2 & 1 & $^{53}$Mn & 46.915 & 47.009  & $^{45}$Ca & 46.323 & 46.406\tabularnewline
\hline 
5/2 & 3 & 0.378 & 46.537 & 46.560  & .174 & 46.149 & 46.280 \tabularnewline
\hline 
3/2 & 3 & 1.290 & 45.625 & 45.695  & 1.435 & 44.888 & 44.991 \tabularnewline
\hline 
11/2 & 3 & 1.441 & 45.474 & 45.454  & 1.554 & 44.769 & 44.763 \tabularnewline
\hline 
9/2 & 3 & 1.620 & 45.295 & 45.309  & - & - & 44.933 \tabularnewline
\hline 
15/2 & 3 & 2.693 & 44.222 & 44.175  & (2.878) & (43.445) & 43.214 \tabularnewline
\hline
\hline 
0 & 0 & $^{54}$Fe & 55.769 & 55.712  & $^{46}$Ca & 56.717 & 56.728, \tabularnewline
\hline 
2 & 2 & 1.408 & 54.360 & 54.286  & 1.346 & 55.371 & 55.501\tabularnewline
\hline 
4 & 2 & 2.538 & 53.230 & 53.307  & 2.575 & 54.142 & 54.332 \tabularnewline
\hline 
6 & 2 & 2.949 & 52.819 & 52.890  & 2.974 & 53.743 & 53.659 \tabularnewline
\hline
\hline 
7/2 & 2 & $^{55}$Co & 60.833 & 60.893  & $^{47}$Ca & 63.993 & 64.014 \tabularnewline
\hline
\hline 
0 & 0 & $^{56}$Ni & 67.998 & 67.950 & $^{48}$Ca & 73.938 & 73.846\tabularnewline
\hline
\end{tabular}
\end{center}
\caption{States in $f_{7/2}$ valence space with spin and seniority listed
in the first and second columns. The $*$ denotes seniority mixed
states in 3B$f_{7/2}$. Following are columns with data for $N=28$
isotones and $Z=20$ isotopes. Three columns for each type of valence
particles list name and excitation energy, experimental binding energy,
and energy from the three-body SM calculation discussed in the text.
All data is in units of MeV. \label{tab:Experimental-data}}

\end{table}

\subsection{Particle-hole symmetry}
The violation of the particle-hole symmetry is due to monopole terms
that are non-liner in the particle-number density.
These terms in the Hamiltonian appear from
three-body and higher rank interactions \cite{Zuker:2003}. For a
single-$j$ and a standard two-body SM the symmetry is exact
and it makes the spectra of $N$ and $\tilde{N}=\Omega-N$ particle systems
identical, apart from a constant shift in energy, here
$\Omega=2j+1$. The
particle-hole conjugation operator $\mathcal{C}$ that acts on a s.p.
state as
$\tilde{a}_{jm}^{\dagger}\equiv\mathcal{C}a_{jm}^{\dagger}\mathcal{C}^{-1}=(-1)^{j-m}a_{j-m},$
transforms an arbitrary $n$-body interaction into itself plus some
Hamiltonian of a lower interaction-rank $H'_{n-1},$ namely $\tilde{H}^{(n)}=(-1)^{n}H^{(n)}+H'_{n-1}.$
The $n=1$ case represents a particles to holes transformation $\tilde{N}=-N+\Omega.$
For the $n=2$ it leads to a monopole shift \begin{equation}
\tilde{H}^{(2)}=H^{(2)}+(\Omega-2N)M,\,\, M=\frac{1}{\Omega}\sum(2L+1)V_{L}^{(2)}.\label{eq:M}\end{equation}
Within a single-$j$ one-body Hamiltonian is a constant of motion, 
being always proportional to $N.$ Thus, following Eq. \eqref{eq:M}, the two-body
interaction is identical for particles
and holes, apart from some constant-of-motion term. 
The interaction of rank 3 and higher violate this
symmetry making excitation spectra of $N$ and $\tilde{N}=\Omega-N$
particle systems different. The experimental data in Tab. \ref{tab:Experimental-data}
shows the particle-hole symmetry violations, for example the excitation
energies of $\nu=2$ states in $N=2$ system are systematically higher
then those in the 6-particle case, indicating a reduced ground state
binding. Using this information a monopole component of the three-body
force can be extracted from the differences in excitation energies
between particle and hole systems, see Fig. \ref{fig:Cumulative} and
discussion below.
\subsection{Seniority}
The $j=7/2$ is the largest single-$j$ shell for which the number
of unpaired nucleons $\nu$, the seniority, is an integral of motion
for any one- and two-body interaction \cite{Schwartz:1954,Talmi:1957}.
Formally, the pair operators $T_{00}^{(2)}$, $T_{00}^{(2)\dagger},$
and particle number $N$ form an SU(2) rotational group, which because
of its analogy to angular momentum is referred to as quasispin. The
relation is established by the operators \[
{\normalcolor
  \mathcal{L}}_{z}=\frac{N}{2}-\frac{\Omega}{4}\,,\quad\mathcal{L}=\frac{\Omega}{4}-\frac{\nu}{2}
\]
with $\mathcal{L}(\mathcal{L}+1)$ being an eigenvalue of the quasispin
vector squared and $\mathcal{L}_{z}$ its magnetic projection.

For a spectrum, the invariance under seniority sets relations between
states of the same  $\mathcal{L}$ but different projection
$\mathcal{L}_{z}$. 
For example, the excitation energies of $\nu=2$ states from
the $\nu=0,$ $0^{+}$ ground state are identical in all even-particle
systems. Using Wigner-Eckart theorem a full set of relations can be
established, see for example sec IIIB in ref. \cite{Volya:2002PRC}
or Ref. \cite{Talmi:1993}. The invariance under quasispin rotations
allows to classify operators in close analogy
to the usual rotations. The s.p. operators associated with
the particle transfer reactions carry $\mathcal{L}=1/2$ and thus
permit seniority change $\Delta\nu=1$. The reactions $^{51}$V($^{3}$He,
d)$^{52}$Cr and $^{43}$Ca(d,p)$^{44}$Ca indicate seniority mixing
as $\nu=4$ final states are populated\cite{Auerbach:1967,Bjerregaard:1967}.
The one-body multipole operators are quasispin scalars for odd angular
momentum, and quasispin vectors for even. Thus, the $M1$ electromagnetic
transitions are given by the quasiscalar operators that do not change
quasispin. The $E2$ operator is a quasivector. In the mid-shell for $^{52}$Cr
and $^{44}$Ca, where $N=\Omega/2$ and $\mathcal{L}_{z}=0$ the $E2$
transitions between states of the same seniority are forbidden. Seniority
can be used to classify the many-body operators $T_{LM}^{(n)}$ and
interaction parameters. The three-body interactions mix seniorities,
one exception is the interaction between $\nu=1$ nucleon triplets
given by the strength $V_{7/2}^{(3)}.$ 

\subsection{Parameter fit and evidence for three-body forces}
To obtain the parameters of the effective Hamiltonian with the three-body
forces we conduct a full least-square $p=$11 parameter fit to $d=$31
data-point (27 in the case of Ca isotopes). The procedure is similar
to a two-body fit outlined for this model space in sec. 3.2 of Ref.
\cite{Lawson:1980}. Schematically ${\bf E=AV}$ where ${\bf E}$
is a set of 31 energies, ${\bf V}$ is a list of 11 interaction parameters
and ${\bf A}$ is 31 by 11 matrix created from the linear form of the Hamiltonian
operator Eq.~\eqref{eq:Hcn}. Due to the seniority
mixing ${\bf A}$ depends on the eigenstates, which in turn are determined
by the interactions ${\bf V};$ thus the overall fitting procedure
is iterative \cite{Brown:2006}. In this example seniority mixing
occurs only in 4 states of the $N=4$ system,  in consequence most of the matrix
elements of ${\bf A}$ are constants. Using the set of experimental
data in Tab. \ref{tab:Experimental-data}, denoted here as ${\bf E}_{ex},$ we determine
${\bf V}=({\bf A}^{T}{\bf A})^{-1}{\bf A}^{T}{\bf E}_{ex}$, where
$T$ and $-1$ superscripts indicate transposed and inverted
matrices. The obtained interaction parameters are used to update non-constant
elements of  ${\bf A}.$ 
The procedure is iterated several times so that the interaction-dependent
components of ${\bf A}$ converge. In Tab. \ref{tab:The-parameters}
the resulting parameters are listed for the $N=28$ proton system, and
for the
neutron $Z=20$ system. The two columns in each case correspond to
fits without (left) and with (right) the three-body forces. The root-mean-square
deviation (RMS) $|{\bf E}_{ex}-{\bf AV}|/\sqrt{d}$ is given for each
fit. The confidence limits given in brackets are inferred from the variances for each
fit parameter  \[
\sigma^{2}(V_{j})=\frac{|{\bf E}_{ex}-{\bf AV}|^{2}}{d-p}({\bf A}^{T}{\bf A})_{jj}^{-1}.\]

\begin{table}
\begin{center}
\begin{tabular}{|c|c|c||c|c|}
\hline 
 & \multicolumn{2}{c||}{$N$=28} & \multicolumn{2}{c|}{$Z$=20}\tabularnewline
\hline
\hline 
 & 2B$f_{7/2}$ & 3B$f_{7/2}$ & 2B$f_{7/2}$ & 3B$f_{7/2}$\tabularnewline
\hline
\hline 
$\epsilon$ & -9827(16)  & -9753(30) & -8542(35)  & -8486.98(72) \tabularnewline
\hline
\hline 
$V_{0}^{(2)}$ & -2033(60)  & -2207(97)  & -2727(122) & -2863(229) \tabularnewline
\hline 
$V_{2}^{(2)}$ & -587(39)  & -661(72)  & -1347(87)  & -1340(176) \tabularnewline
\hline 
$V_{4}^{(2)}$ & 443(25)  & 348(50) & -164(49)  & -198(130) \tabularnewline
\hline 
$V_{6}^{(2)}$ & 887(20) & 849(38) & 411(43) & 327(98) \tabularnewline
\hline
\hline 
$V_{7/2}^{(3)}$ &  & 55(28)  &  & 53(70) \tabularnewline
\hline 
$V_{5/2}^{(3)}$ &  & -18(70)  &  & 2(185) \tabularnewline
\hline 
$V_{3/2}^{(3)}$ &  & -128(88)  &  & -559(273) \tabularnewline
\hline 
$V_{11/2}^{(3)}$ &  & 102(43)  &  &  51(130) \tabularnewline
\hline 
$V_{9/2}^{(3)}$ &  & 122(41) &  & 272(98) \tabularnewline
\hline 
$V_{15/2}^{(3)}$ &  & -53(29) &  & -24(73)\tabularnewline
\hline
\hline 
RMS & 120 & 80 & 220 & 170\tabularnewline
\hline
\end{tabular}
\end{center}
\caption{Interaction parameters of 2B$f_{7/2}$ and 3B$f_{7/2}$ SM Hamiltonians
determined with the least-square fit are given in keV.\label{tab:The-parameters} }

\end{table}

The reduction of the RMS deviation, for example for $Z=28$ isotones
it drops from 120 keV to about 80 keV, is not the only evidence in support
of the three-body forces.  The fit parameters are stable within quoted
error-bars even if some questionable data-points are removed. The
energies from the three-body fit listed in Tab. \ref{tab:Experimental-data}.
are comparable or even better
than the results from many two-body SM calculations
in the expanded model space \cite{Lips:1970,Poves:2001}. However,
such comparisons are difficult since different models have different
number of parameters and were fit to different sets of nuclei.

In Tab. \ref{tab:BE} we discuss the renormalization of pairing by
considering a minimal fit limited to the ground
states and a single three-body term. The fit is similar to Ref. \cite{Talmi:1957}, but has
a seniority conserving three-body force given by the $\nu=1$ triplet
operator $T_{jm}^{(3)}\sim a_{jm}^{\dagger}T_{00}^{(2)}$ with the
strength $V_{7/2}^{(3)}$. This interaction is equivalent to a density-dependent pairing
force \cite{Zelevinsky:2006}. In a single-$j$ shell the renormalization of
pairing by a particle-number dependent strength 
\begin{equation}
V_{0}^{(2)'}=V_{0}^{(2)}+\Omega\frac{N-2}{\Omega-2}V_{j}^{(3)}
\end{equation}
allows for an exact treatment of the three-body term.
The ground state
energies with $\nu=0$ or 1 are
\begin{equation}
E=\epsilon
N+\frac{N-\nu}{\Omega-2}\left((\Omega-N-\nu)\frac{V_{0}^{(2)'}}{2}+(N-2+\nu)M'\right),\label{eq:Egs}\end{equation}
which is a usual expression \cite{Talmi:1957,Volya:2002PRC}, but
includes a renormalized pairing strength denoted with prime. 
The
results from the minimal fit are shown in Tab. \ref{tab:BE}, they are
consistent with the full fit in Tab. \ref{tab:The-parameters}.

\begin{table}
\begin{center}
\begin{tabular}{|c|c|c||c|c|}
\hline 
 & \multicolumn{2}{c||}{$N=28$} & \multicolumn{2}{c|}{$Z=20$}\tabularnewline
\hline 
$\epsilon$ & -9703(40) & -9692(40) & -8423(51) & -8403(55)\tabularnewline
\hline 
$V_{0}^{(2)}$ & -2354(80) & -2409(110) & -3006(120) & -3105(156)\tabularnewline
\hline 
$M$ & 1196(40) & 1166(50) & -823(55) & -876(76)\tabularnewline
\hline 
$V_{7/2}^{(3)}$ & - & 18(20) & - &  31(31)\tabularnewline
\hline
RMS & 50 & 46 & 73 & 65\tabularnewline
\hline
\end{tabular}
\end{center}
\caption{Interaction parameters for the minimal $f_{7/2}$ SM determined with
the linear least-squared fit of 8 binding energies. In brackets the
variances for each parameter are shown. The two columns for isotopes
and isotones are fits without and with the three-body term. \label{tab:BE}}

\end{table}

\begin{figure}
\begin{center}
\includegraphics[width=3.5in]{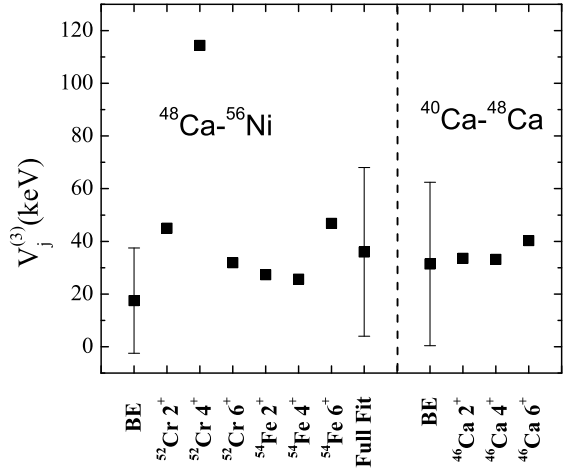}
\end{center}
\caption{Cumulative data on $V_{7/2}^{(3)}$ seniority $\nu=1$ effective three-body
force in $N=48$ isotones, left, and $Z=40$ isotopes to the right.
The point labeled as BE comes from a fit to 8 binding
energies in Tab.\ref{tab:BE} and includes a fitting error-bars. The
point labeled as {}``Full Fit'' corresponds to a fit of all 31 levels
in $N=28$ isotones with 6 parameters for s.p.
energy, two-body force and $V_{7/2}^{(3)}$. Other individual points correspond
to extraction of $V_{7/2}^{(3)}$ from excitation energy, always compared
to $N=2$ system ($^{50}$Ti or $^{42}$Ca) \label{fig:Cumulative}}

\end{figure}

In Fig. \ref{fig:Cumulative} we give a cumulative picture showing the
$V_{7/2}^{(3)}$ term found with different methods. As discussed above, due to
the particle-hole symmetry and
seniority conservation, excitation energies of $\nu=2$:
$2^{+},\:4^{+},$ and $6^{+}$
states in $N=2,4,$ and 6 -particle systems should be identical. The
$V_{7/2}^{(3)}$ coefficient can be found assuming that it is responsible
for most of the mass difference. For example, the difference in excitation
energies of these states between $^{50}$Ti and $^{52}$Cr equals
to $8V_{7/2}^{(3)}/3$. The independent result on $V_{7/2}^{(3)}$
inferred from these observations, the binding energy fit,
and the fit to all states in the $N=28$ isotones with 4 parameters
are summarized
in Fig. \ref{fig:Cumulative}. The point that corresponds to the 
$4^{+}$ state in $^{52}$Cr in Fig. \ref{fig:Cumulative} is not in agreement with
the rest of the data, it demonstrates the seniority mixing discussed below. 

It follows from Tabs. \ref{tab:The-parameters} and \ref{tab:BE}, and
Fig. \ref{fig:Cumulative} that
within the error-bars
the three-body interaction is isospin invariant; it is the same for proton and
neutron valence spaces.

\subsection{Seniority mixing in  $^{52}$Cr}
\begin{figure}
\begin{center}
\includegraphics[width=2.5in]{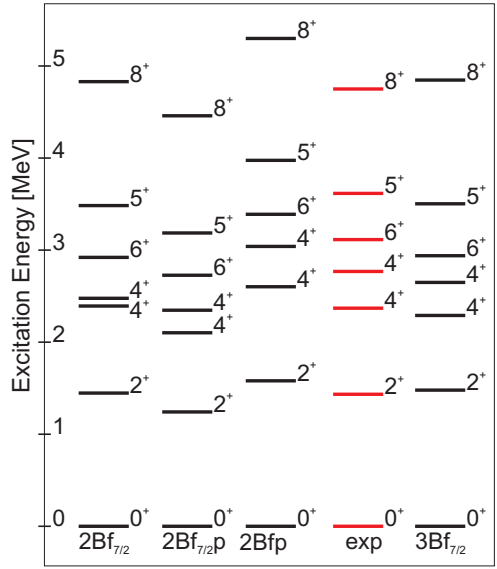}
\end{center}
\caption{Spectrum of $52$Cr.\label{fig:Spectrum-of-52Cr}}

\end{figure}

The mid-shell case of $^{52}$Cr, see Fig.\ref{fig:Spectrum-of-52Cr},
is interesting to discuss. Here, in addition to 2B$f_{7/2}$
and 3B$f_{7/2}$ interactions from Tab. \ref{tab:The-parameters}
we perform a large scale SM calculation 2B$f_{7/2}p$ (includes $p_{1/2}$
and $p_{3/2}$) and 2B$fp$ (entire $fp$-shell, truncated to $10^{7}$
projected m-scheme states) using FPBP two-body SM Hamiltonian \cite{Brown:2001}.
Similar results in a more restrictive valence space can be found in
Ref. \cite{Lips:1970}. %

The level repulsion between neighboring $4_{1}^{+}$ and $4_{2}^{+}$
states is generated by the seniority mixing, the observed energy difference
of 400 keV is not reproduced by the 2B$f_{7/2}$ (84 keV) model. As
seen in Fig. \ref{fig:Spectrum-of-52Cr} the discrepancy remains in
the extended two-body model 2B$f_{7/2}p$ (200 keV). Although,
the full 2B$fp$ model reproduces the splitting, the excessive intruder
admixtures over-bind the ground state and effectively push 
all states up in excitation energy. The 3B$f_{7/2}$
model is in good agreement with experiment; its predictions for
the seniority mixing are $\nu(4_{1}^{+})=2.82$ and $\nu(4_{2}^{+})=2.71,$
as calculated
from the expectation value of the pair operator $\langle
T_{00}^{(2)^{\dagger}}T_{00}^{(2)}\rangle=(N-\nu)(2j+3-N-\nu)/(4j+2).$
The $2_{1}^{+}$ state is relatively pure $\nu(2_{1}^{+})=2.006$. 

The seniority mixing violates quasispin selection rules \cite{Armstrong:1965,Bjerregaard:1967,McCullen:1964,Monahan:1968,Pellegrini:1973}
which in the past have been explained by the two-body models beyond
the single-$j$ \cite{Auerbach:1967,Engeland:1966,Ginocchio:1963,Lips:1970,McCullen:1964},
however such models not always describe all of the features observed
in experiment. In particular, to explain electromagnetic transitions
sizable variations of effective charges are needed \cite{Brown:1974}
and the particle transfer spectroscopic factors do not show large amount
of strength outside the $f_{7/2}$ valence space \cite{Armstrong:1965}.
In Tab. \ref{tab:B(E2)} $B(E2)$ transitions rates from all models
are compared to experiment. To make a fair comparison
the combination of the nuclear radial overlap and effective charge is
normalized using observed $E2$ rate for the transition $2_{1}\rightarrow0_{1}$
in the 2B$f_{7/2}$, 2B$f_{7/2}p,$ and 2B$fp$ models. The parameter
for the 3B$f_{7/2}$ model is identical to the one used in the 2B$f_{7/2}.$
The small difference in $2_{1}\rightarrow0_{1}$ $B(E2)$ between
the 3B$f_{7/2}$ and 2B$f_{7/2}$ models is a result of the $\nu=4$ admixture
in the $2_{1}^{+}$ state. The strong $\nu=4$ and $\nu=2$ seniority
mixing between $4_{1}^{+}$ and $4_{2}^{+}$ states impacts forbidden
transitions; for example, $E2$ transitions $4_{2}\rightarrow2_{1}$
and $6_{1}\rightarrow4_{2}$ become allowed. 

\begin{table}
\begin{center}
\begin{tabular}{|c|c|c|c|c|c|}
\hline 
 & 2B$f_{7/2}$ & 2B$f_{7/2}p$ & 2B$fp$ & 3B$f_{7/2}$ & Experiment\tabularnewline
\hline
\hline 
$2_{1}\rightarrow0_{1}$$^{(*)}$ & 118.0 & 118.0 & 118 & 117.5 & 118$\pm35$\tabularnewline
\hline 
$4_{1}\rightarrow2_{1}$ & 130.4 & 122.5 & 105.8 & 73.2 & 83$\pm15$$^{(1,2)}$\tabularnewline
\hline 
$4_{2}\rightarrow2_{1}$ & 0 & 3.3 & 15.1 & 56.8 & 69$\pm18$\tabularnewline
\hline 
$4_{2}\rightarrow4_{1}$ & 125.2 & 59.3 & 2.6 & 0.5 & \tabularnewline
\hline 
$2_{2}\rightarrow0_{1}$ & 0 & 0.003 & 0.9 & 0.5 & 0.06$\pm0.05$\tabularnewline
\hline 
$2_{2}\rightarrow2_{1}$ & 119.2 & 102.2 & 101.9 & 117.1 & 150$\pm35$\tabularnewline
\hline 
$2_{2}\rightarrow4_{1}$ & 0 & 10.8 & 34.4 & 19.9 & \tabularnewline
\hline 
$2_{2}\rightarrow4_{2}$ & 57.8 & 7.2 & 5.2 & 38.7 & \tabularnewline
\hline 
$6_{1}\rightarrow4_{1}$ & 108.9 & 86.2 & 56.3 & 57.8 & 59$\pm20^{(1)}$\tabularnewline
\hline
$6_{1}\rightarrow4_{2}$ & 0 & 9.3 & 27.6 & 51.1 & 30$\pm10^{(1)}$\tabularnewline
\hline
\end{tabular}
\end{center}
\caption{B(E2) transition summary on $^{52}$Cr expressed in units $e^{2}$fm$^{4}$.
The data is taken from \cite{Huo:2007}. $^{(*)}$In the 2B $f_{7/2}p$
and 2B$fp$ models we use $0.5$(neutron) and $1.5$(proton) effective
charges, the overall radial scaling is fixed by the B(E2,$2_{1}\rightarrow0_{1}$).
$^{(1)}$The life-time error-bars were used. $^{(2)}$There are conflicting
results on life-time; we use DSAM (HI, x$n\gamma$) data from Ref.
\cite{Huo:2007}, which is consistent with \cite{Brown:1974}.\label{tab:B(E2)}}
\end{table}

The proton removal spectroscopic factors in Tab. \ref{tab:Proton-removal-spectrascopic}
show a similar picture, where the seniority mixing has a strong impact
on transitions. In support of the three-body forces as a source of
the mixing it was argued in Ref. \cite{Armstrong:1965} that the sum
of spectroscopic factors for $4^{+}$ states is close to 4/3 which
is consistent with the observation in Ref. \cite{Armstrong:1965} and does not
support the expanded valence space where spectroscopic factors are
reduced due to fragmentation of the single-particle strength.
\begin{table}
\begin{center}
\begin{tabular}{|c|c|c|c|c|c|}
\hline 
 & 2B$f_{7/2}$ & 2B$f_{7/2}p$ & 2B$fp$  & 3B$f_{7/2}$ & Exp\tabularnewline
\hline
\hline 
$0{}_{1}^{+}$ & 4.00 & 3.73 & 3.40 & 4.00 & 4.00\tabularnewline
\hline 
$2{}_{1}^{+}$ & 1.33 & 1.14 & 0.94 & 1.33 & 1.08\tabularnewline
\hline 
$4{}_{1}^{+}$ & 0.00 & 0.13 & 0.34 & 0.63 & 0.51\tabularnewline
\hline 
$4{}_{2}^{+}$ & 1.33 & 1.11 & 0.70 & 0.71 & 0.81\tabularnewline
\hline 
$6{}_{1}^{+}$ & 1.33 & 1.28 & 1.28 & 1.33 & 1.31\tabularnewline
\hline
\end{tabular}
\end{center}
\caption{Proton removal spectroscopic factors. The experimental data is taken
from $^{51}$V($^{3}$He,d)$^{52}$Cr reaction \cite{Armstrong:1965}.
Within error-bars this data is consistent with
results \cite{Huo:2007}. \label{tab:Proton-removal-spectrascopic}}
\end{table}

\section{Three-body forces in oxygen isotopes}
The above single-$j$ example is remarkable due to its transparency and
simplicity. The general SM case, however, is complicated
by an enormously large number of parameters and thus difficulty of
the fit  \cite{Poves:1981, Hees:1989, Hees:1990,Zuker:2003}. 
Selecting dynamically relevant components of the many-body forces
requires an in-depth microscopic understanding of their origin.
Establishment of the physically
relevant set of the operator basis $T_{LM}^{(n)^{\dagger}}(\alpha)$ is
an important start. 
As discussed in the introduction, for $n>2$ the index $\alpha$ must 
include an additional information about the coupling scheme, the choice of which is not unique.
Previous ideas on selecting the best set of 
triplet operators include a possibility of using the $\nu=1$
operators for each single-particle level \cite{Zelevinsky:2006}. For
$j=7/2$ the three-body force associated with this
operator, discussed in Fig. \ref{fig:Cumulative}, is indeed a
dominating component in binding.  
However, it is not clear if such construction, built upon s.p. levels,
is the best choice in a general case given renormalization of the s.p. effective degrees of
freedom by the two-body interaction. The two-body Hamiltonian of the
pairing type would, for instance, suggest the use of quasiparticles. 
 
Perusing this idea we propose an alternative approach which 
assumes a hierarchy of forces, where higher rank components of the
Hamiltonian are perturbative, and the operator basis are selected using 
the many-body dynamics.

Consider $H_{n-1}$, $n\ge 3$ Hamiltonian to be determined by some procedure. While building a higher rank
forces $H_{n}=H_{n-1}+H^{(n)}$, we assume $H^{(n)}$ to be perturbatively small.
Thus, within the lowest order perturbation theory the $n$-particle 
wave-functions of $H_{n-1}$ and $H_{n}$ are the same and can be found 
by diagonalizing $H_{n-1}$
\[H_{n-1}|n;LM\alpha\rangle=E_{n;L}(\alpha)\,|n;LM\alpha\rangle.\]
We use these eigenstates to define a full set of
$n$-body operators $T_{LM}^{(n)}(\alpha)$  as
$|n;LM\alpha\rangle=T_{LM}^{(n)^{\dagger}}(\alpha)|0\rangle,$ which we
view as the most relevant basis. With 
a perturbative nature in mind the term $H^{(n)}$ is diagonal in these basis, 
$V_{L}^{(n)}(\alpha\beta)=0$ if $\alpha\ne\beta.$ Thus, the number of
parameters is reduced. 
Further steps can be taken to discuss the
significance of the diagonal parameters. When pairing is
important one can take only those states (basis operators) that correspond to the lowest
quasiparticle excitations. Experimental data can be used for guidance.
For example, if the $n$-particle states are known and identified 
experimentally to have energies $E_{L}^{(exp)}(n;\alpha),$ a direct
fit can be done by setting the corresponding $n$-body
interaction parameters as
$V_{L}^{(n)}(\alpha,\alpha)=E_{n;L}^{(exp)}(\alpha)-E_{n;L}(\alpha),$
so that the new Hamiltonian reproduces exactly the experimental
energies. 

There are some issues to stress. 
Certainly, the transition from $n=1$ to $n=2$, is
not a subject to this approach. One has to have a starting SM Hamiltonian
$H_{SM}^{(2)}$ determined from G-matrix techniques or by other methods,
see \cite{Brown:2001} and references therein. It is possible
to rewrite the two-body Hamiltonian as a diagonal structure by
introducing new pair operators, this is useful for perturbative
adjustments of the two-body
interactions.  

The two-body SM Hamiltonian can be used as a primary component of
interaction, defining many-body operators, and treating all higher rank forces as perturbations. 
It is important for this approach to stay within the perturbation
theory. Departing a perturbative form, it is feasible with this construction to create a
Hamiltonian that exactly reproduces energies of all states within a given
valence space, tests show that in this case the many-body forces have an
inverse hierarchy with higher rank ones giving a bigger contribution.

It is an established practice in the SM approach to include a mass dependence
of the two-body forces. For a short range delta-type interaction
the radial overlap integrals scale as $R^{-3/2}$, where $R$ is the
radius of the nucleus. Thus, in terms of the mass-number $A$ the
two-body interaction $H^{(2)}\sim A^{-1/2}.$ At the opposite extreme
the long-range Coulomb leads to an $A^{-1/6}$ scaling. The fits to
experimental data lead to a compromising middle value $A^{-0.3}$\cite{Brown:2006}.
The many-body forces are expected to be short range, requiring
all participating particles to be localized. The resulting scaling
that follows from this argument is 
\begin{equation}
H^{(n)}\sim A^{(1-n)/2}\,,{\rm so}
\quad
H^{(n)}(A)=\left(\frac{A_{c}+n}{A}\right)^{(n-1)/2}H^{(n)}(A_{c}),
\label{eq:scaling}
\end{equation}
where $A_c$ is the mass of the core. 
At this stage it is not clear if this argument is valid and if scaling
should be included. 

\begin{figure}
\includegraphics[width=4.5in]{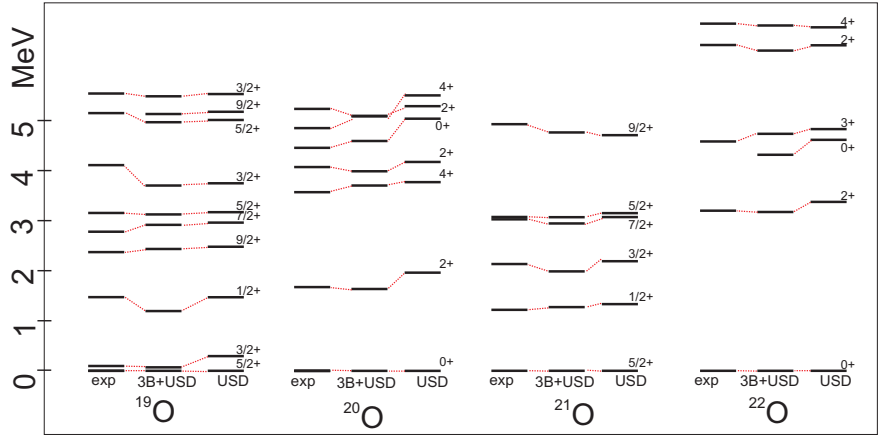}
\caption{Spectrum of $^{19-22}$O isotopes. For every nucleus,
  experimentally observed states are compared with the spectrum that
  includes three-body forces and with the two-body USD SM Hamiltonian,
  from left to right, as identified at the bottom.  
\label{fig:USD}}
\end{figure}

As a demonstration we discuss here a 3-body force in the case of
oxygen isotopes. For the two-body interaction Hamiltonian we take a
USD shell model \cite{USD}. The total number of triplet operators, not
counting magnetic projections, is 37 which in a general 3-body
interaction Hamiltonian gives a large number of parameters. 
Examination of experimental data for  $^{19}$O and results from
the different shell model Hamiltonians USD, USDA and USDB \cite{Brown:2006} show a rather
systematic difference; in particular for the lowest
$5/2^+_1$, $3/2^+_1$ and $1/2^+_1$ states. These are one
quasiparticle excitations. Thereby, we define the
corresponding triplet operators $T_{jm}^{(3)^{\dagger}}$ with
$j=5/2,\,3/2,$ and 1/2 from the three-particle eigenstates of the USD Hamiltonian; and 
specify the three-body 
interaction in the diagonal form 
\begin{equation}
H^{(3)}=\sum_{j=5/2,3/2,1/2}\,V_{j}^{(3)}\sum_{m=-j}^{j}\,
T_{jm}^{(n)^{\dagger}}\, T_{jm}^{(n)}.
\label{eq:3b}
\end{equation}
For Fig. \ref{fig:USD} we fit the three parameters in Eq. \eqref{eq:3b} to
the ground states in even systems and to the three lowest states with one
unpaired particle in the odd systems for mass $A=19$ to 22 oxygen
isotopes. The values from the best fit are $V^{(3)}_{5/2}=45$ keV,
$V^{(3)}_{3/2}=-179$ keV, and  $V^{(3)}_{3/2}=-231$ keV. 

The improvement in the spectrum, seen in  Fig. \ref{fig:USD}
is significant. Certainly, this first study is to be continued, there
is a possibility to examine more interaction terms, discuss
scaling of the matrix elements, and to consider fitting all parameters
for one-, two-, and three-body components together. Modifying
perturbatively the two-body part should not invalidate the quality of
the USD-defined three-body basis.  

\vspace {-1.2 cm}
\section{Conclusion}

Dealing with many-body forces, understanding their
origins, structure, and hierarchy of renormalizations is an important
component for a successful solution of a many-body problem. This
presentations aims to continue the discussion in Ref. \cite{Eisenstein:1973,
Poves:1981, Hees:1989, Hees:1990, Zuker:2003, Volya:2008, Volya:arxiv}
related to the 
phenomenological three-body forces within the context of the
nuclear shell model approach. 
The study of nuclei in the $0f_{7/2}$ shell shows evidence of such forces
through an overall fit to data with full examination of uncertainties, via
examination of binding energies and associated differences in
excitation spectra, and with an in-depth analysis of violations of
symmetries in the structure of wave functions.

The general SM problem with many-body forces is complicated by 
a large number of
parameters, the absence of a good microscopic approach, difficulties in
fits and questions related to
renormalizations of strengths. 
These issues are discussed and some
methods for dealing with them are proposed. In particular, 
in analogy to a Hartree-Fock procedure where single-particle states
are defined in the way to best represent the dynamics of the system, we
propose here methods to identify the most relevant many-body
operators.   These techniques are
demonstrated using a chain of oxygen isotopes.

Support from the U. S. Department of Energy, grant DE-FG02-92ER40750 is acknowledged. 

%\bibliographystyle{h-physrev}
%\bibliography{n-body}

\end{document}